# Design of Experiments aided Holistic Testing of Cyber-Physical Energy Systems

Arjen A. van der Meer*, Cornelius Steinbrink†, Kai Heussen‡, Daniel E. Morales Bondy‡, Merkebu Z. Degefa§,
Filip Pröstl Andrén¶, Thomas I. Strasser¶, Sebastian Lehnhoff†, Peter Palensky*

*Delft University of Technology, Delft, The Netherlands, {a.a.vandermeer, p.palensky}@tudelft.nl
†OFFIS – Institute for Information Technology, Oldenburg, Germany {steinbrink, lehnhoff}@offis.de
‡Technical University of Denmark, Lyngby, Denmark {kh, bondy}@elektro.dtu.dk
§SINTEF Energi AS, Trondheim, Norway {merkebuzenebe.degefa}@sintef.no
¶AIT Austrian Institute of Technology, Vienna, Austria {filip.proestl-andren, thomas.strasser}@ait.ac.at

*Abstract*—The complex and often safety-critical nature of cyber-physical energy systems makes validation a key challenge in facilitating the energy transition, especially when it comes to the testing on system level. Reliable and reproducible validation experiments can be guided by the concept of design of experiments, which is, however, so far not fully adopted by researchers. This paper suggests a structured guideline for design of experiments application within the holistic testing procedure suggested by the European ERIGrid project. In this paper, a general workflow as well as a practical example are provided with the aim to give domain experts a basic understanding of design of experiments compliant testing.

## I. INTRODUCTION

Validation and testing are commonly named as important milestones in the roll-out of cyber-physical systems in general and cyber-physical energy systems (CPES) in particular [1], [2]. There are different opinions on how to realize such validation: Sha *et al.* [3], for example, advocate splitting complex systems into subsystems that can be formally validated. While formal checking of subsystems is critical, it does not provide guarantees that interactions across subsystem in the complex CPES will not develop undesired behaviors. In other words, system-wide testing is necessary [2]. Since CPES are typically too complex and heterogeneous to be addressed by formal validation methods, an experimental approach is required, be it software-based, hardware-based, or a combination of both in a so-called Hardware-in-the-Loop (HIL) manner [4].

Research institutions that regularly conduct CPES validation experiments have set up hardware laboratories, real-time software and simulation environments to do so. However, due to heterogeneous testing setups and workflows, the replication of experimental results is typically challenging and thus often neglected. In order to foster reproducibility and exchange between test infrastructures, the ERIGrid project has set out to develop a standardized process for development and documentation of validation experiments [5], [6]. Due to its system-wide and multi-domain focus, the process is a so called *holistic testing procedure*. The process has been developed within ERIGrid and has been adopted by trans-national access user projects [7] and other European research and development projects like ELECTRA IRP [8], SmartNet [9], etc. It provides a structured workflow with various concept definitions so that researchers can use a common vocabulary when talking about purpose, objects, domains and other aspects of their test case. The approach supports hardware, software, simulation, and mixed experiment types. However, in its current state, the process is focused on the transfer of test setups among different testing environments. In order to have comparable and reproducible results, their analysis must be robust against statistical fluctuations. Such a goal is commonly achieved by employing a set of tools known as *Design of Experiments* (DoE) [10].

DoE is a technique aiming to clarify the cause and effect relationships of factors and outputs while reducing the cost of doing the experiment. It comprises a number of statistical tools and concepts suitable for dealing with variance and fluctuation in experiment inputs, setups, and results. In its origin, the DoE approach has been devised for the realization of physical experiments [11], but with the advent of more complex simulation experiments similar principles have been applied to software-based testing [12].

This paper extends the current state of the ERIGrid holistic testing procedure by proposing a practical workflow for DoE usage. In the holistic testing procedure, DoE maintains its essential purpose by checking the contributions of the various complex factors and defining the needed redundancies for meaningful statistical evaluation of the outputs. The purpose of this study is to identify relevant DoE practices for designing holistic CPES related experiments. Furthermore, the experiments should utilize the capabilities of different lab infrastructures.

The paper is structured as follows: In Section II the holistic testing procedure is described with its current state and limitations. Section III provides an overview of the basis of DoE and its concepts relevant for CPES testing followed by Section IV where an exemplary simulation experiment is established to demonstrate the proposed workflow. Section V, finally, concludes the paper and provides an outlook about planned future work.

## II. HOLISTIC TESTING APPROACH

The *holistic testing procedure* defined by ERIGrid outlines an approach to realize CPES experiments in such way that the purpose and context of the tests are separated from the specific lab setups realizing the experiments [5], [13]. By following this method, the purpose of the test is explicit and can then be mapped to and replicated across different test environments.

The method is represented graphically in Fig. 1, which shows the stages of a holistic test description (details can be found in [14].). First, a *Test Case* must be formulated, which is a description of a test on similar description level as a Use Case (a well-known concept for system development [15]; refer to *system* or *technical use case* level). At this step the *Object under Investigation* (OuI) is defined within a *System under Test* (SuT) implementing Functions under Test (FuT), with a specific *Purpose of Investigation* (PoI) and associated *Test Criteria*. Three overall PoI categories have been identified: (i) *characterization*, (ii) *validation*, and (iii) *verification*. The SuT identifies domains, system components and relevant hosted functions, and their interactions, which affect the Test Criteria and must therefore be taken into consideration. Test Criteria may address any quantifiable aspect of the SuT. For validation purposes these can be motivated by key performance indicators given the use case for the respective function under investigation. In other words, the Test Case defines the reason for the test, what to test, and what to test for.

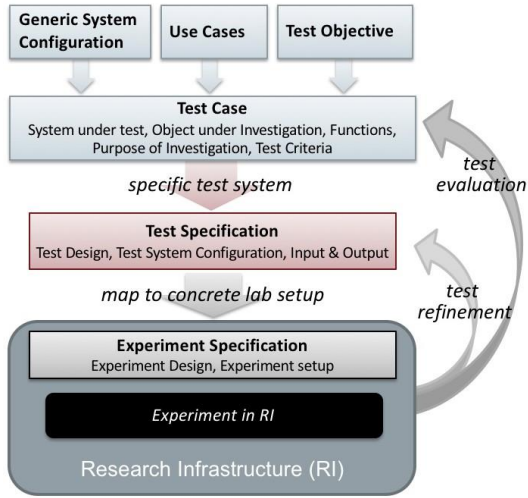

Fig. 1. The holistic test description stages separate the purpose of the tests from the laboratory setup used for testing [13].

The *Test Specification* addresses a specific aspect of the Test Case, where the Test System Configuration is a specific instance of the SuT (i.e., defining a specific grid topology, component configurations, and variables of the SuT and OuI). Furthermore, it draws boundaries on the test, defining inputs and outputs, and a *Test Design*. The Test Design explains the planning of how the test is to be carried out to facilitate the acquisition of relevant qualification data.

As several variants of a Test System can be relevant for a single Test Case, a Test Case may contain several independently assessed Test Criteria, and limitations of available test environments can often be anticipated, it is recommended to formulate a Qualification Strategy which explains how a testing need is divided among several Test Specifications.

In the *Experiment Specification*, the Test Specification is mapped to the available components and functionality of the lab where the experiment is executed. This consists of two main parts: (i) the *Experiment Setup*, which reflects the Test System, and (ii) the *Experiment Design*. The later must plan in detail and document all the needed information for the execution, evaluation, and replication of the test.

Test Case, Test Specifications, and Experiment Specifications are to be recorded via dedicated templates (provided in [14]) to facilitate information exchange between researchers. The focus of this work is to illustrate how the field of DoE is to be applied to the Test and Experiment designs.

## III. DESIGN OF EXPERIMENTS FOR CPES TESTING

Several publications offer both broad and specialized introductions to the topic of DoE (e.g., [10], [12], [16], [17], [18]), introducing the statistical methods of DoE and their application in detail. This paper will focus on providing a general understanding of the terms and concepts of DoE which are relevant for the holistic testing procedure.

In order to avoid confusion of the terminology, terms that are generally used in DoE are marked with $^x$ while terms of the ERIGrid holistic testing procedure are marked with $^o$. Terms used in both fields are marked with $^{x/o}$.

### A. Basic Terms and Concepts of DoE

One hurdle to take for researchers that are new to DoE is the terminology. In the following, we attempt to provide a compact yet sufficient summary of the terms that form the basis of the DoE vocabulary. As mentioned above, experiment parameters are called *factors*$^x$ in DoE. Their different values that are tested in experiments are called *levels*$^x$. A set of parameter values for one test run (one level chosen for each factor) is called a *treatment*$^x$ and a set of treatments is chosen following a given *design*$^{o/x}$. An integral part of the DoE workflow is the choice of a design that supports the purpose of investigation and is compatible with the number of factors and levels proposed for the experiment. In general, the Purpose of Investigation$^o$, and the associated Test Criteria$^o$, is associated in some way with identifying the effect of factor values on the output of the system at hand, which may also be called *response*$^x$ or *target metrics*$^{o/x}$. These are the basic terms of DoE, but they do not suffice to get a complete picture of the field. On the one hand, the experiment output has to be subjected to statistical analysis methods in order for DoE to show its full potential. A variety of methods exist so that naming all of them would exceed the scope of this paper. Next to the analysis, a number of additional concepts are important for DoE, but can hardly be summarized under one umbrella term. Such concepts of which DoE users need to be aware are, e. g., *randomization*[x], *blocking*[x] or *confounding*[x]. The meaning of this concepts is explained further below in the context of their application in CPES testing. Further explanations of all basic DoE terms and concepts can be found in various text books [16], [17].

In more in-depth DoE literature the term *factor*[x] typically describes all aspects that may influence an experiment's outcome. In this context it is important to differentiate between two classes of factors. The first class is called *treatment factors*[x]. The experiment has been designed to obtain information about the influence of these factors, which makes them explicitly important. The second class is called *nuisance factors*[x]. These factors do not lie in the focus of the experiment, but they could not have been removed either. Therefore, they need to be considered to make sure that their influence is not wrongly attributed to the treatment factors. Note that the differentiation between treatment and nuisance factors is not done in terms of controllability. Both types of factors can be either controllable of uncontrollable. Treatment factors that are fully controllable may also be called *experimental factors*[x] while uncontrollable ones may oftentimes at least be classified and are thus called *classification factors*[x] [17]. As an example, the testing of a Photovoltaic (PV) panel control unit can be considered. The influence of weather may be of explicit interest so that it is considered a treatment factor. It cannot be fully controlled, but different classifications may be assigned like "cloudy"/"sunny" or "morning"/"afternoon" and so forth.

In general, there are certain items to be considered in choosing the appropriate DoE techniques for the right problem. The first is the viable *number of experiments* which itself depends on the time required for a single experiment. The second is limiting *the number of factors* in order to reduce the size of the problem and the effort required to solve it. The third is the careful choice of *the number of levels* to have a reasonable number of experiments while allowing a good interpolation on the design space. The fourth, as mentioned above, is to clarify *the aim* of the experiment in order to determine the suitable DoE techniques (design and analysis methods) with sufficient outputs. The following section illustrates the planning of these steps within the structure of the holistic testing procedure.

*B. Structured DoE via Holistic Testing Approach*

One major development goal of the holistic testing process is to provide a structured framework for integration of DoE into the workflow of CPES testing. The key to this is the multi-stage approach of test and experiment specification. In the holistic testing procedure, practitioners start out by defining the SuT and PoI of their study and only then proceed to specify their input parameters and sources of uncertainty, which can be translated into treatment factors and nuisance factors. This way, the most important information (goal of the study and number of factors) is clear in an early stage of the study and can be used to make informed decisions about target metrics, the number of levels, experimental design and analysis methods. These are to be picked mainly in the stage of Experiment Specification. Another important aspect of the holistic multi-stage approach is the concept of refinement and re-evaluation. Practitioners are encouraged to use preliminary experiment results to improve and refine test and experiment specifications in an iterative way. This fosters the employment of good practices in DoE like screening experiments: if at first the Test Specification is defined with a large number of factors, a DoE approach is chosen that is focused on identifying the impact of the factors (typically with few levels per factor). This *sensitivity analysis* allows pointing out negligible factors that are then excluded from the updated Test Specification. Finally, a truly holistic testing workflow requires DoE since statistically sound testing forms the basis of comparability of results between research infrastructures and thereby the overall reproducibility of CPES tests.

Fig. 2 displays a suggested workflow that structures a DoE process (right side) within the holistic testing procedure (left side). As described above, the holistic testing procedure guides the practitioner in obtaining the information from their application case that is needed to conduct DoE. Thus, many specifications in the holistic process directly inform DoE choices (e.g., number and types of factors, number of levels). The Test Case description, on the other hand, provides more general information that has to be aggregated and considered by the practitioner to make justifiable choices for analysis methods and experimental designs. These choices again have to be documented in the templates of the holistic process.

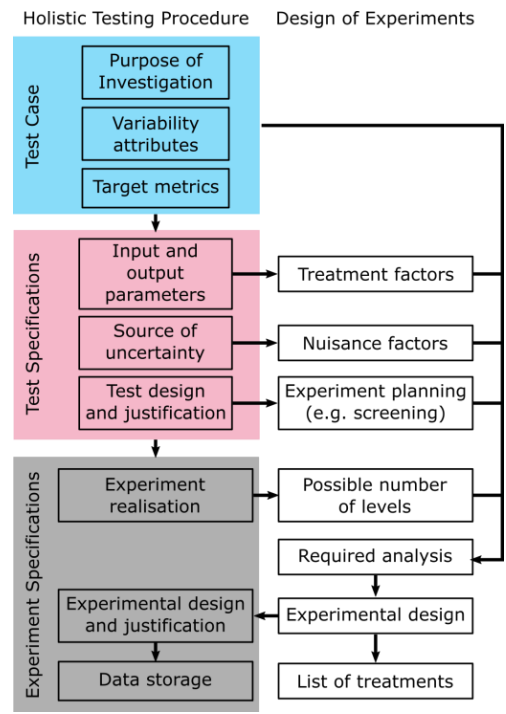

Fig. 2. DoE as a part of the holistic testing workflow.

The presented workflow is one aid to help practitioners of the holistic testing procedure to integrate DoE into their CPES experiments. Aside from that, it is important that they are enabled to make informed decisions regarding their choice of designs and analysis methods. Unfortunately, the fields of both, statistics and CPES, are too diverse to allow for formulation of standard application cases. Nevertheless, the following sections provide a general overview of the applicability of selected DoE concepts within CPES testing.

*1) Choice of analysis methods:* As mentioned above, an experimental design (leading to a set of treatments) and the employed analysis methods should be chosen in accordance to one another. Furthermore, these decisions have to be informed by the aim of the experiment, the number of factors and further considerations of the system under test. However, this work addresses some of these concepts rather independently. This is done to enable a stronger focus on some considerations that are specific to CPES validation and the holistic testing procedure. The choice of analysis methods is for now discussed in the context of the typical goals of CPES experiments.

Considering the aim of an experiment, researchers must differentiate between preliminary goals on the one hand and the actual PoI on the other hand. A typical preliminary goal is reduction of the number of considered factors via sensitivity analysis. As mentioned before, the associated experiments are typically called screening. The actual PoI of the testing experiment is defined in the context of the holistic Test Case.

The holistic procedure differentiates between *characterization$^o$*, *validation$^o$*, and *verification$^o$* as PoI$^o$. These different purposes may be translated into different statistical problems. For a characterization purpose, the test practitioner will typically focus on establishing polynomial description models that approximate the mathematical connection between experiment factors and outputs (polynomial regression). Verification, on the other hand, can be seen as a hypothesis test since researchers are typically testing against a reference. In such a context DoE is typically used to establish an *analysis of variance$^x$* (ANOVA) and choose acceptable risks for type I and type II errors—the incorrect rejection of a true null hypothesis (type I) and the acceptance of an incorrect null hypothesis (type II). Validation can be seen as a mixed case between verification and characterization: it involves testing against references like the former, but requires an amount of result interpretation like the latter since the references are typically not well-defined. As a consequence, validation experiments might require practitioners to use DoE to establish description models as well as ANOVA. Incidentally, ANOVA as well as description models may also be needed for screening experiments dependent on the screening purpose. The aforementioned screening for sensitivity analysis typically requires ANOVA. On the other hand, screening may also be used to check for nonlinearities in the effect of certain factors on the output. In this case DoE can be used to check for the plausibility of different regressions.

All of the aforementioned general attributions between PoI and analysis methods are summarized in Table I.

*2) Choice of experimental design:* The choice of an experimental design depends on the aspired analysis method and on the properties of the SuT, namely number of factors, expected nonlinearities and realizable number of treatments as outlined above. Experimental designs can be roughly separated into two categories: (i) classical designs, and (ii) modern, simulation-oriented designs [19]. Classical designs are strongly focused on providing as much information as possible with a strongly limited number of treatments since an experiment run is typically considered to be associated with a high effort or cost. As a consequence, classical designs often make simplifying assumptions about the SuT, such as considering only small numbers of factors or linear system behavior (two-level designs). Modern designs, on the other hand, typically assume higher numbers of treatments to be possible. Accordingly, interactions of higher numbers of factors can be analyzed. Furthermore, designs may be space-filling so that nonlinear system behavior may be analyzed in more detail. However, since modern designs have been mostly established with software experiments in mind, they are typically associated with test systems that are strictly deterministic and thus do not display fluctuations. Since classical designs have been established for stochastic systems, they place a stronger focus on reproduction of factor values.

TABLE I
ANALYSIS METHODS ACCORDING TO EXPERIMENTAL PURPOSE.

|  |  | Major analysis tool | Comment |
|---|---|---|---|
| Preliminary purpose | Screening (SA) | ANOVA | Many factors, few levels |
|  | Nonlinearity checking | Description model (regression) | Few factors, many levels |
| Major PoI | Characterization | Description model (regression) | Analysis tools are guideline suggestions |
|  | Validation | Regression + ANOVA |  |
|  | Verification | ANOVA |  |

All popular designs have in common that they avoid *confounding* of factors. Confounding means in this context that the possible effects of two or more factors cannot be distinguished from one another in the statistical analysis. The occurrence of confounding is (with some exceptions) a sign of bad experimental design and cannot be fixed with statistical analysis so that more experimentation is necessary. As a consequence, choosing any one established experimental design is better than choosing none at all. The best design choice is highly dependent on the application case at hand. Table II gives a rough overview of the basic characteristics of the two major classes of designs to help guide practitioners into the most suitable direction. For a more refined choice, special literature on the particular categories of designs should be consulted. Note that this work avoids the common mapping which associates classical DoE with hardware and modern DoE with software experiments. While this may be a reasonable distinction in some domains, it cannot be unconditionally applied to the CPES domain. In fact, CPES experiments exist in a spectrum between pure hardware and software testing. Automated or hybrid HIL setups may be configurable to easily process large numbers of treatments while very complex simulation setups, on the other hand, may be limited to few treatments due to restricted computation capacity. All in all, the choice of an experimental design can be translated to finding an individual trade-off between an acceptable number of treatments and a desired amount of information.

TABLE II
EXPERIMENTAL DESIGN OVERVIEW.

| Design category | Characteristics | Examples |
| --- | --- | --- |
| Classical designs | Small number of treatments, focus on fluctuations, trade-off between number of factors and number of levels (interactions vs nonlinearities) | Full-factorial, Fractional-factorial, Plackett-Burman, Central-Composite, Box-Behnken |
| Modern designs | Large number of treatments, fluctuations mostly neglected, large number of factors, space-filling designs | Latin-Hypercube, Sobol sequence, Monte Carlo, Orthogonal arrays |

*3) Handling of different factor types:* As mentioned above, factors can be divided into different categories. While experimental designs provide a structured way of working with treatment factors, nuisance factors typically need some special consideration to prevent them from distorting the result analysis. More precisely, nuisance factors exist in different degrees of controllability. In worst case, a nuisance factor is completely unknown (and thus also uncontrollable). The common way to mitigate the effect of such a factor is *randomization*. That means that the treatments chosen for an experiment are conducted in a randomized order. This may allow for identification of underlying effects upon experiment reproduction. Since the risk of unidentified nuisance factors is always given, randomization should be always conducted.

In best case, nuisance factors are known and controllable. They are commonly handled via *blocking*, where a set of levels is defined for the factor in question. Then, an experimental design for the treatment factors is established for each of these levels ("blocks") with treatment reproduction between the blocks and randomization within each blocks. This allows for distinguishing block effects from treatment factor effects.

TABLE III
HANDLING OF NUISANCE FACTORS.

| Factor type | Handling concept |
| --- | --- |
| Unknown | Randomization |
| Known, controllable | Blocking |
| Known, uncontrollable | ANCOVA |

Nuisance factors that are known but fully uncontrollable, have to be handled within the experiment analysis, typically via *analysis of covariance* (ANCOVA). It allows for analysis of the effects of treatment factors while statistically controlling the effects of the aforementioned nuisance factors. Table III summarizes the types of nuisance factors with the corresponding concepts for handling them. More detailed explanation of these concepts and their application can be found e.g., in [17].

IV. EXAMPLE USE CASE

*A. Test Case Description*

As an example of how the DoE methods can be applied for holistic testing of CPES, a use case on fault ride through (FRT) of a wind power plant is considered. It is common practice to have FRT capabilities as a part of the grid connection requirements that the plant owner must obey. Usually, these capabilities are confined in time versus voltage profiles. Practically, the wind turbines within the plant must stay connected during voltage sags as long as the voltage at the Point of Common Coupling (PCC) stays above the required profile and shall recover the pre-fault active power output according to a minimum recovery rate afterwards. During and after FRT it is manifest to maintain short term voltage and frequency stability of the overall system. The power electronic interface and favorable controllability of wind turbine generators (WTG) can be used to support the stability.

The PoI of this test case is to verify the ability of the wind power plant to support stability and to characterize under which conditions the stability can be boosted. The SuT together with the DoE workflow is shown in Fig. 3. The SuT consists of the power system, exhibiting the generator speed deviation ($\Delta\omega_{G1}$) and voltage amplitude $|U|_{PCC}$ responses as functions under investigation (i.e., the dependent variables), the WTG converter, yielding the interface between the converter controls and the grid as OuI, and the mechanical part of the WTG as a boundary. The main control functions impacting the voltage and frequency behavior are the voltage control and active power control through maximum power point tracking. The former aims to support the voltage through additional reactive current injection (aRCI) during and after faults, indicated by a proportional gain factor $K_{aRCI}$. The latter regulates post-FRT active power recovery rate by a maximum rate limit on $i_d^*$ (i.e., $R_p$).

The bottom half of Fig. 3 shows how DoE can be applied. The main factors of interest to the response variables of the SuT are the control parameters described above and system variables such as generator dispatching, the pre-fault, and the retained voltage profile at the PCC (i.e., caused by the fault location). The DoE process consists of a screening part and the actual design part. For the sake of brevity we will focus on the former, thereby applying ANOVA.

In-field FRT experiments are quite complicated. First, igniting faults for stability analysis will not be appreciated by the majority of connected partners and mistakes in the experimental setup can cause quite some real damage and corresponding cascaded effects. Secondly, voltages, loads, generator dispatch, and wind power fluctuate so conditions differ continually. This makes it hard to consider these as treatment factors.

When assessing the test criteria by simulation studies instead, a significant flexibility gain is achieved: virtually every factor can be regarded controllable and hence a treatment factor. This gives rise to the application of (deterministic) simulation experiments as an integral part of DoE-aided holistic testing. To limit the eventual set of treatments we will first perform a screening on $K_{aRCI}$, the current limiting strategy, and $R_p$. After qualitatively assessing their sensitivity on the response variables it can be decided in which way the factors shall be blocked at a limited number of level or regarded as treatment variables in the actual design.

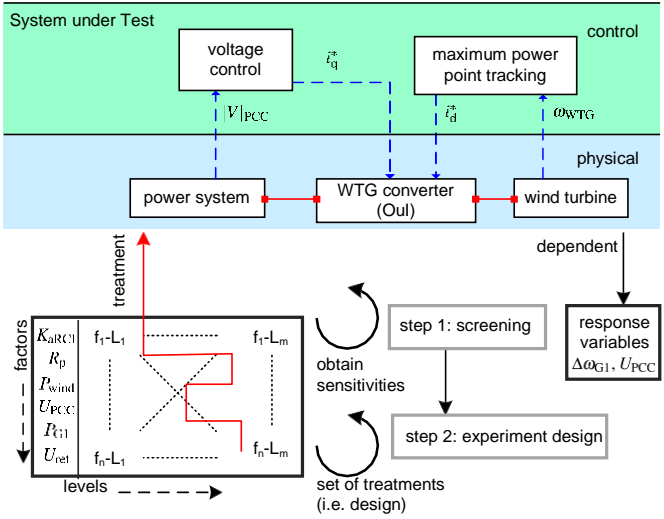

Fig. 3. Overview of DoE methodology and its relation to the functions and system under test of the example case study.

The SuT is split into two parts, each of which is included into a separate simulator, after which the overall response is obtained via co-simulation. The power system is the IEEE 9-bus test system [20], G3 of which will be replaced by proxy model of a wind power plant (i.e., the static generator in PowerFactory). The wind power plant itself is represented aggregately by a MATLAB/Simulink dynamic WTG converter model rated 120 MVA. This model takes as an input from PowerFactory $U_{PCC}$ and $P_{PCC}$ and provides the static generator model with d- and q-axis projections of the reference currents (i.e., $\vec{i}_d^*$ and $\vec{i}_q^*$). The co-simulation is set up according to the functional mockup interface (FMI) standard and run using the FMI++ package [21]. PowerFactory is exported according to the FMI for co-simulation specification and the WTG model is encapsulated as a functional mockup unit (FMU) by FMI for model exchange 2.0, the solver of which is provided by the FMI++ platform. Both FMUs internally apply adaptive time-stepping whereas the synchronization interval is fixed at 1 ms.

### B. Screening Simulation Results

First we will study the effect of the voltage-dependent additional reactive current injection on the system voltage and frequency deviation. For the sake of simplicity the rotor speed deviation of G1 is taken as a measure for rotor angle stability. The proportional control gain $K_{aRCI}$ is varied from 0 (no effect at all) to 2.0, meaning that a 50% retained voltage yields injection of rated reactive current (i.e., 120 MVar). The results are shown in Fig. 4. It can be seen that the impact on the terminal voltage of the WTG is relatively small: despite the retained voltage during FRT being slightly elevated for higher values of $K_{aRCI}$, the voltage response is dominated by excitation systems of the remaining synchronous machines. For the frequency response things are different. It can be observed that for $K_{aRCI} \geq 1.0$ the amplitude of $\Delta\omega_{G1}$ is clearly attenuated. A higher $K_{aRCI}$ hence supports the rotor angle stability.

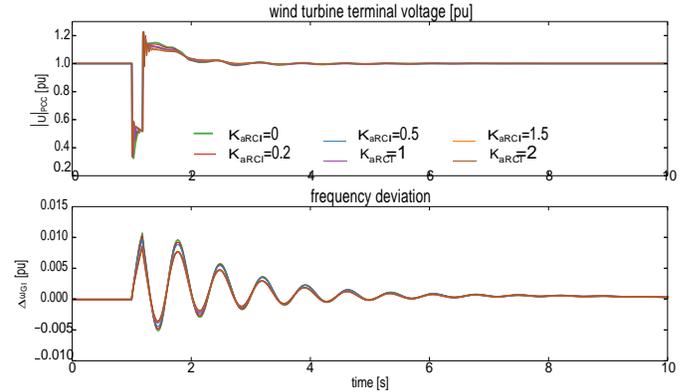

Fig. 4. The effect of additional reactive current injection on frequency and voltage behavior in the system under test.

Next, the effect of the current limiting strategy of the WTG converter is addressed. During faults the magnitude of the current set points exceed the current rating of the front-end converter, i.e., $\sqrt{\vec{i}_d^{*2} + \vec{i}_q^{*2}} \geq |I|_{lim}$, the current needs to be curtailed along either the active d-axis or the reactive q-axis. Taking $K_{aRCI} = 1.0$, the effect of both limiting strategies on $\Delta\omega_{G1}$ can be seen in Figure 5. The merit of prioritizing $\vec{i}_q^*$ is clear and this factor will hence be blocked for further analysis.

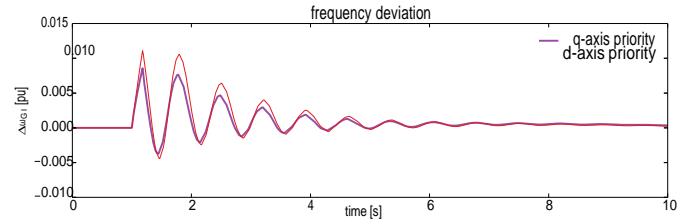

Fig. 5. The effect of active or reactive current priority for current limiting on frequency and voltage behavior in the system under test.

Finally, the active power recovery rates $R_p$ are considered as screening factors. During faults, the output power is reduced to 0 in favor of reactive power support, and hence needs to recover after fault clearance. A higher $R_p$ implies faster recovery of pre-fault WTG power output. The results are depicted in Fig. 6, the scaling of which has been slightly altered to visualize the effects of low ramping rates. It is manifest that mainly the frequency is impacted by this parameter.

This preliminary screening allows us to limit the amount of treatment factors for the experiment design. For $K_{aRCI}$ both $U_{PCC}$ and $\Delta\omega_{G1}$ are influenced. As a matter of fact, it is expected that the impact on the voltage response becomes more significant with remote faults causing shallow voltage dips, which was not considered during the screening phase. $K_{aRCI}$ thus needs to be fully considered as a treatment factor. For the limiting strategy it was clear that q-axis prioritization is favorable for the frequency response as compared to the alternative. This factor can be blocked for further analysis. As for $R_p$, values higher than 10 pu/s can be considered technically infeasible because of WTG protection, whereas very low values impair the frequency response considerably and even cause WTGs to interfere with the governing systems of the conventional plants. Hence, the results show that although this parameter must be considered as a treatment factor, its levels and range in the set of treatments can be limited.

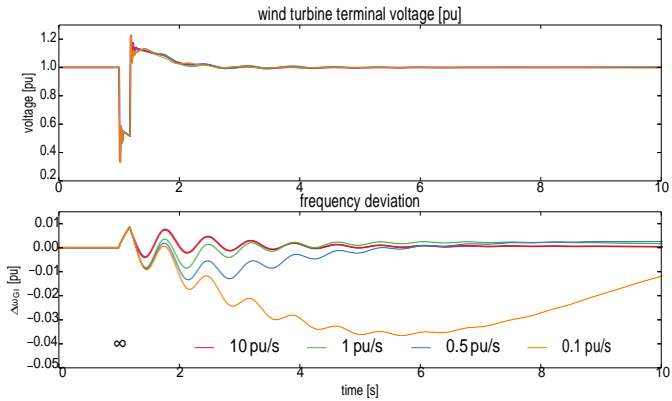

Fig. 6. The effect of active power recovery rates on voltage and frequency response of the system under test.

## V. Conclusions

The paper at hand aims at improving the state of European-wide validation and testing of CPES. It does so by illustrating the advancement of the ERIGrid holistic testing procedure via the explicit consideration of DoE concepts. This work suggests a DoE workflow that encourages CPES researchers to conduct statistically sound experiments within a structured, well-documented holistic process. Furthermore, general guidelines have been established that function as starting points for the systematic integration of DoE concepts into CPES testing. Additionally, a time-domain simulation-based CPES test case boasts a practical example for DoE-compliant research.

As an outlook on future work, ERIGrid will continue its work on improvement of international CPES validation. Consequently, more guideline material is to be expected to consolidate application of DoE in the CPES domain. Further improvement of the guidelines will, e. g., include a discussion of threats to validity. These threats may diminish knowledge gained about interdependency between factors and target metrics ("internal validity") or transferability of the experiment results ("external validity"). Some threats are already mitigated through the employment of concepts like randomization or blocking. For other threat definitions, however, active debate is going on regarding their applicability to lab-based, non-human testing [22] so that appropriate guidelines are valuable for the CPES community.

## Acknowledgment

This work is supported by the European Communitys Horizon 2020 Program (H2020/2014-2020) under project "ERIGrid" (Grant Agreement No. 654113).

## References

[1] J. Wan, H. Yan, H. Suo, and F. Li, "Advances in cyber-physical systems research," *KSII Transactions on Internet and Information Systems*, vol. 5, no. 1, pp. 891–1908, 2011.
[2] T. H. Morris *et al.*, "Engineering future cyber-physical energy systems: Challenges, research needs, and roadmap," in *North American Power Symposium (NAPS)*, 2009.
[3] L. Sha, S. Goapalakrishnan, X. Liu, and Q. Wang, "Cyber-physical systems: a new frontier," in *2008 IEEE International Conference on Sensor Networks, Ubiquitous, and Trustworthy Computing*, 2008.
[4] T. Strasser, F. Pröstl Andren, G. Lauss, R. Bründlinger *et al.*, "Towards holistic power distribution system validation and testing – an overview and discussion of different possibilities," *e & i Elektrotechnik und Informationstechnik*, vol. 134, no. 1, pp. 71–77, 2017.
[5] A. A. van der Meer, *et al.*, "Cyber-Physical Energy Systems Modeling, Test Specification, and Co-Simulation Based Testing," in *2017 Workshop on Modeling and Simulation of Cyber-Physical Energy Systems (MSCPES)*, 2017.
[6] T. Strasser, C. Moyo, R. Bründlinger, S. Lehnhoff *et al.*, "An integrated research infrastructure for validating cyber-physical energy systems," in *Industrial Applications of Holonic and Multi-Agent Systems*. Springer International Publishing, 2017, pp. 157–170.
[7] ERIGrid, *Trans-national Access*. Available: https://erigrid.eu/transnational-access/.
[8] L. Martini, H. Brunner, E. Rodriguez, C. Caerts *et al.*, "Grid of the future and the need for a decentralised control architecture: the web-of-cells concept," *CIRED – Open Access Proceedings Journal*, vol. 2017, no. 1, pp. 1162–1166, 2017.
[9] G. Migliavacca, M. Rossi, D. Six, M. Dzamaria *et al.*, "SmartNet: A H2020 project analysing TSO-DSO interaction to enable ancillary services provision from distribution networks," *CIRED – Open Access Proceedings Journal*, vol. 2017, no. 1, pp. 1998–2002, 2017.
[10] M. J. Anderson, P. J. Whitcomb, *Design of experiments*. Wiley Online Library, 2000.
[11] R. A. Fisher, *Statistical methods for research workers*. Genesis Publishing Pvt Ltd, 1925.
[12] J. P. C. Kleijnen, *Design and analysis of simulation experiments*. vol. 20. New York: Springer, 2008.
[13] M. Blank, S. Lehnhoff, K. Heussen, D. E. Morales Bondy *et al.*, "Towards a Foundation for Holistic Power System Validation and Testing," in *21th IEEE International Conference on Emerging Technologies and Factory Automation (ETFA'2016)*, 2016.
[14] K. Heussen, D. E. Morales Bondy, (editors), *Smart Grid configuration validation scenario description method*. ERIGrid Consortium, 2017. Available: https://erigrid.eu/dissemination/.
[15] International Electrotechnical Comission, *IEC 62559-2:2015 - Use Case Methodology*. Available: https://webstore.iec.ch/publication/22349.
[16] M. Cavazzuti, Design of experiments. *Optimization methods*. Springer Berlin Heidelberg, 2013. 13-42.
[17] D. C. Montgomery, *Design and analysis of experiments*. John Wiley & Sons, 2017.
[18] L. W. Condra, *Reliability improvement with design of experiment*. CRC Press, 2001.
[19] A. A. Giunta, S. F. Wojtkiewicz Jr., and M. S. Eldred, "Overview of modern design of experiments methods for computational simulations," in *41st Aerospace Sciences Meeting and Exhibit*, 2003.
[20] P. M. Anderson and A. A. Fouad, *Power System Control and Stability*, 1st ed. Ames, IA: The Iowa State University Press, 1977.
[21] The FMI++ Library, Available online at http://sourceforge.net/projects/fmipp/, Last accessed: Feb. 2018.
[22] J. Siegmund, N. Siegmund and S. Apel, "Views on Internal and External Validity in Empirical Software Engineering," in *2015 IEEE/ACM 37th IEEE International Conference on Software Engineering (ICSE)*, 2015.